\documentclass[12pt,a4paper]{article}

\usepackage{a4}

\usepackage{amsfonts}
\usepackage{amssymb}
\usepackage{graphicx}
\usepackage{psfrag}
\usepackage{dsfont}

\usepackage[sc,small]{caption2}
\setcaptionwidth{14cm}

\newcommand{\beqn}{\begin{eqnarray}}
\newcommand{\eeqn}{\end{eqnarray}}
\newcommand{\be}{\begin{equation}}
\newcommand{\ee}{\end{equation}}
\newcommand{\non}{\nonumber \\}
\newcommand{\mbb}{\mathbb}

\newcommand{\ce}{{\cal E}}

\newcommand{\cn}{{\cal N}}

\newcommand{\cv}{{\cal V}}

\newcommand{\co}{{\cal O}}

\newskip\humongous \humongous=0pt plus 1000pt minus 1000pt

\newif\ifdtup

\begin{document}

\title{}
\author{}
\date{}
\thispagestyle{empty}

\begin{flushright}
\vspace{-3cm}
{\small MIT-CTP-3675\\
        NSF-KITP-2005-56 \\[-.1cm]
        hep-th/0508171}
\end{flushright}
\vspace{1cm}

\begin{center}
{\bf\LARGE
On Volume Stabilization \\[.5cm]
by Quantum Corrections}

\vspace{1.5cm}

{\bf Marcus Berg}$^{\dag}$
{\bf,\hspace{.2cm} Michael Haack}$^{\dag}$
{\bf\hspace{.1cm} and\hspace{.2cm} Boris K\"ors}$^{*}$
\vspace{1cm}

{\it
$^{\dag}$Kavli Institute for Theoretical Physics, University of California \\
Santa Barbara, California 93106-4030, USA\\

$^*$Center for Theoretical Physics, Laboratory for Nuclear Science \\
and Department of Physics, Massachusetts Institute of Technology \\
Cambridge, Massachusetts 02139, USA \\

$^*$II. Institut f\"ur Theoretische Physik der Universit\"at Hamburg \\
Luruper Chaussee 149, D-22761 Hamburg, Germany\\

$^*$Zentrum f\"ur Mathematische Physik, Universit\"at Hamburg \\
Bundesstrasse 55, D-20146 Hamburg, Germany
}

\vspace{1cm}

{\bf Abstract}
\end{center}
\vspace{-.5cm}

We discuss prospects for stabilizing the volume modulus of $\cn=1$
supersymmetric type IIB orientifold compactifications using only
perturbative corrections to the K\"ahler potential. Concretely, we
consider the known string loop corrections and tree-level
$\alpha'$ corrections. They break the no-scale structure of the
potential, which otherwise prohibits stabilizing the volume
modulus. We argue that when combined, these corrections provide
enough flexibility to stabilize the volume of the internal space
without non-perturbative effects, although we are not able to
present a completely explicit example within the limited set of
currently
available models. Furthermore, a certain amount of fine-tuning is
needed to obtain a minimum at large volume.

\clearpage


\section{Introduction}

In a companion paper \cite{gg1} we computed one-loop corrections
to the K\"ahler potentials of certain supersymmetric orientifolds
of type IIB string theory. Here, we interpret these results in the
context of volume stabilization. The two papers can be read
independently.

To make contact with phenomenology, a viable string theoretical
model must deal with the problem of stabilizing the scalar moduli
fields. One way to generate a potential for many of these scalars
is by turning on background fluxes. In type IIB orientifolds,
fluxes lead to {\it no-scale} potentials, which stabilize some
moduli but leave the K\"ahler moduli unfixed, among them the
overall volume of the internal space \cite{Giddings:2001yu}. The
general form of this flux-induced potential is given by the
standard F-term expression of $\cn=1$ supergravity\footnote{We
always refer to IIB orientifolds with D3- and D7-branes here. In
models with D5- and D9-branes, D-term potentials appear as well
\cite{Grimm:2004uq}.}
\beqn\label{scalpot}
V = e^K \left[ K^{M\bar N} D_M W D_{\bar N} \bar W -3 |W|^2
\right] \ ,
\eeqn
where the superpotential only depends on the complex structure
moduli $U^\alpha$ and the axio-dilaton $S$,
\beqn\label{fluxsupo}
W_{\rm flux} = W(S,U^\alpha)\ .
\eeqn
Here $M,\, N$ run over all the fields (chiral multiplets) of the
theory. The K\"ahler covariant derivative is defined as usual,
$D_M W = (\partial_M + K_M) W$, and subscripts indicate
derivatives.

For a IIB string compactification on a Calabi-Yau manifold
with a single K\"ahler modulus $\rho$ whose imaginary part
measures the overall volume of the internal space, the classical
K\"ahler potential for $\rho$ has the form\footnote{We leave out
additive constants in writing the K\"ahler potential, such as
$-3\ln[-i]$ in this case, since they do not affect the K\"ahler
metric, but we do take care of these factors in the prefactor
$\exp(K)$ of the potential (\ref{scalpot}).}
\beqn\label{KKLT}
K (\rho,\bar\rho) = -3 \ln (\rho-\bar \rho)\ .
\eeqn
The expression is ``classical'' in the sense that it is leading
order in string perturbation theory and in the $\alpha'$
expansion of the effective action. In particular, the moduli space
factorizes and the K\"ahler metric is block-diagonal with one
factor for the K\"ahler moduli. The superpotential
(\ref{fluxsupo}) and the K\"ahler potential (\ref{KKLT}) together
satisfy
\beqn\label{noscale}
K^{\rho \bar\rho} D_{\rho} W D_{\bar\rho} \bar W = 3|W|^2\ ,
\eeqn
which leads to the no-scale property \cite{Cremmer:1983bf} of the
classical potential of orientifold flux compactifications. This
generalizes to several K\"ahler moduli $\rho^{\alpha}$ and to any
perturbatively generated superpotential, since the real parts (RR
scalars) of the K\"ahler moduli obey a shift symmetry that forbids
their appearance in the superpotential to all orders in
perturbation theory \cite{Witten:1985bz,Giddings:2001yu}. The
resulting scalar potential fails to produce masses for the
K\"ahler moduli, and thus leads to the problem of volume
stabilization at the level of the leading perturbative
approximation.

One is  then led to go beyond the leading approximation, to break
the no-scale structure of the potential, thereby generating a
potential also for the K\"ahler moduli. There are several ways to
do this. The first is to allow the superpotential to depend on the
K\"ahler moduli. This approach was pursued by Kachru, Kallosh,
Linde and Trivedi (KKLT) in
\cite{Kachru:2003aw}. Due to the Re$(\rho^\alpha)$ shift symmetries  mentioned above, the construction of
\cite{Kachru:2003aw} has to rely on non-perturbative strong coupling
effects on the world volume of D-branes (gaugino condensation on
D7-branes or D3-brane instantons) to produce a
superpotential that depends
on the K\"ahler moduli. This leads to a scalar
potential with enough structure to stabilize also K\"ahler moduli.
Of course, such non-perturbative effects are hard to control.
Furthermore, some fine-tuning of the various input parameters is
required to produce satisfactory minima at large volume.

The second way to break no-scale structure does not rely on
non-perturbative strong coupling phenomena. The K\"ahler potential
(\ref{KKLT}) naturally receives corrections in perturbation
theory, both from $\alpha'$ corrections and from string loops.
These corrections break the factorization of the total moduli
space (into K\"ahler moduli times complex structure and
axio-dilaton) and lead to more complicated dependence of the
K\"ahler potential on the K\"ahler moduli. As a consequence, the
no-scale structure of the potential is lifted, and it is no longer
impossible to stabilize the volume modulus simply by perturbative
corrections to the K\"ahler potential, i.e.\ without any
dependence on the volume in the superpotential.\footnote{An
alternative approach was considered in \cite{Balasubramanian:2004uy,Balasubramanian:2005zx,Conlon:2005ki}, combining the effects of
a non-perturbative superpotential with non-trivial dependence on
$\rho$ and those of the $\alpha'$ corrections. The result reported there was that the perturbative
corrections cannot be neglected and that a class of
non-supersymmetric vacua with large volume can be found.
Also, \cite{Antoniadis:2002gw} considered volume stabilization
by perturbative string corrections in nonsupersymmetric models.}

For $\alpha'$ corrections, the breaking of no-scale structure was
demonstrated in \cite{Becker:2002nn}. Unfortunately, the
corrections to the no-scale potential from purely sphere-level
$\alpha'$ corrections are not sufficiently well understood to
claim stabilization of the volume modulus. Here, we will add
type IIB orientifold one-loop corrections that we determined in
\cite{gg1} for certain models. We will see that they constitute
another source of no-scale breaking and introduce additional
structure sufficient to stabilize the volume modulus, at least in
principle.\footnote{This was also discussed in
\cite{vonGersdorff:2005bf}; see \cite{Bobkov:2004cy} as well.} As it turns out, in the very limited
set of examples analyzed in \cite{gg1} a significant amount of
fine-tuning is required to stabilize the volume large enough so
that one can neglect further corrections. Nevertheless, we
consider the results promising, in that one can now look for
concrete models where the volume modulus can be stabilized by
perturbative corrections alone.


\section{General remarks on the large volume expansion}
\label{secgeneral}

In terms of string perturbation theory, the no-scale K\"ahler
potential (\ref{KKLT}) for the volume modulus $\rho$ arises from
sphere diagrams, expanded to leading order in $\alpha'$. It
receives a host of corrections beginning already at disk level,
i.e.\ tree-level of the open string. At disk level, it is known
that the argument of the logarithm in (\ref{KKLT}) is shifted by a
function of the open string scalars that we collectively denote
$A$ and on the complex structure moduli $U$, cf.\
\cite{Antoniadis:1996vw,Aldazabal:1998mr,DeWolfe:2002nn,Jockers:2004yj},
in the form
\beqn
K(\rho,\bar\rho)&=&-\ln [\rho-\bar \rho] \nonumber \\
\stackrel{{\rm disk}}{\longrightarrow} \quad
K(\rho,\bar\rho,A,\bar A, U, \bar U)&=&-\ln [\rho-\bar\rho +
f(A,\bar A, U, \bar U)]\ ,
\eeqn
This can be considered the classical tree-level K\"ahler potential
at leading order in $\alpha'$.

Beyond leading order in $\alpha'$, there is a correction known at
order $(\rho-\bar\rho)^{-3/2}$
\cite{Becker:2002nn} whose coefficient is proportional to the
Euler number of the internal Calabi-Yau mani\-fold. Beyond leading
order in the string coupling, the corrections in \cite{gg1} are
one-loop corrections from Klein bottle, annulus, and M\"obius
strip diagrams, that are typically suppressed by
$(\rho-\bar\rho)^{-1}$ and $(\rho-\bar\rho)^{-2}$, with
coefficients that depend on the complex structure and axio-dilaton
as well as on open string moduli. All of these corrections to
(\ref{KKLT}) ruin the factorization of the moduli space (the
K\"ahler moduli space is no longer separate), and break the
no-scale structure.

Putting these pieces together, the expansion of the K\"ahler
potential involves terms of the form\footnote{See also
\cite{Berglund:2005dm,Giddings:2005ff} for related discussions.}
\beqn \label{Kgeneral}
K &=& -3 \ln \Big[\rho-\bar \rho + f_1(A, \bar A, U, \bar U)\Big]
+ \frac{1}{\rho - \bar \rho} \Big[ \frac{f_2(A, \bar A, U, \bar U)}{S-\bar S} + \ldots \Big]\\
&& + \frac{1}{(\rho - \bar \rho)^{3/2}} \Big[ \alpha (S -\bar S)^{3/2} + \ldots
\Big]
+ \frac{1}{(\rho - \bar \rho)^2} \Big[ f_3(A, \bar A, U, \bar U) +
\ldots \Big] + \ldots\ , \nonumber
\eeqn
where we omitted $\rho$-independent terms and those that are more
suppressed in the large volume limit. The dots in the brackets
involve terms with more inverse powers of the dilaton field
$S-\bar S$. These arise from corrections above one-loop, that are
suppressed at small string coupling. Without fluxes, there are no
further $\alpha'$ corrections at sphere level. But with fluxes,
there could be terms leading in the dilaton expansion compared to
the terms shown in (\ref{Kgeneral}). For example, the term
suppressed by $(\rho -\bar\rho)^{-2}$ might receive corrections
already at sphere level, from a term proportional to $G_3^2 R^3$
(with $G_3$ the 3-form flux). In addition, in the presence of
D-branes, there may be $\alpha'$ corrections from disk diagrams
that could contribute to (\ref{Kgeneral}). Both of these potential
additional $\alpha'$ corrections are poorly understood and we will
limit our discussion of $\alpha'$ corrections to the
$(\rho-\bar\rho)^{-3/2}$ term in (\ref{Kgeneral}).

We now consider a superpotential of the form (\ref{fluxsupo}) and
``integrate out'' $S$ and $U$ by setting $D_S W = D_U W = 0$. This
fixes $S$ and $U$ at $S^{(0)} +\co((\rho-\bar\rho)^{-1})$ and
$U^{(0)} +\co((\rho- \bar \rho)^{-1})$, with constants $S^{(0)}$
and $U^{(0)}$, and we further assume that this is a minimum of the
potential with respect to $S$ and $U$. The idea behind this
hierarchal integrating-out is an assumption of a separation of
scales, where the complex structure and the axio-dilaton receive
masses much larger than the typical scale of the potential for
$\rho$ and $A$. Substituting the K\"ahler potential of
(\ref{Kgeneral}) into (\ref{scalpot}), the scalar potential takes
the form
\beqn
V = e^{K} \Big( K^{\bar \rho \rho} K_{\bar \rho} K_{\rho} +
K^{\bar \rho A} K_{\bar \rho} K_{A} + K^{\bar A \rho} K_{\bar A}
K_{\rho} + K^{\bar A A} K_{\bar A} K_{A} - 3 \Big) |W|^2 \ .
\eeqn
The breaking of the no-scale structure is manifest whenever the
factor in parenthesis no longer vanishes. However, computing the
potential using (\ref{Kgeneral}) one finds that there is still no
term of order $(\rho -\bar\rho)^{-1}$ in the large-volume
expansion, even though there is such a term in the one-loop
correction to the K\"ahler potential. Instead, the leading term
that breaks no-scale arises from the $\alpha'$ correction in
(\ref{Kgeneral}). Therefore, for a flux-induced superpotential of
the form (\ref{fluxsupo}) the correction to the scalar potential
is stronger suppressed at large volume than one could naively have
expected from (\ref{Kgeneral}). It would be interesting to
determine the effects of the K\"ahler potential (\ref{Kgeneral})
together with a non-perturbative superpotential with non-trivial
$\rho$-dependence, along the lines of
\cite{Balasubramanian:2005zx,Conlon:2005ki}.

With these observations the scalar potential that follows from
setting $D_S W = D_U W = 0$, but $D_\rho W \neq 0\neq D_A W$ (so
that supersymmetry is broken spontaneously), becomes
\be \label{V}
V = \frac{1}{(-i(\rho-\bar \rho))^3} \Bigg[ \frac{c_1}{(-i(\rho -
\bar
\rho))^{3/2}} + \frac{c_2}{(-i(\rho - \bar \rho))^{2}} +\ldots
\Bigg]|W|^2\ .
\ee
The prefactor comes from $\exp(K)$, and the coefficients $c_1$ and
$c_2$ are functions of $A$, and depend on the values of the
constants $S^{(0)}$ and $U^{(0)}$ that, in turn, depend on the
flux values. Note that to determine higher terms in the expansion
in the inverse volume, one would also have to solve the relations
$D_S W = D_U W = 0$ for $S$ and $U$ to the next order in
$(\rho-\bar \rho)^{-1}$, and substitute into the scalar potential.

To find a minimum of the potential that stabilizes the volume, one
should now minimize this potential with respect to $A$ and $\rho$.
As noted also in \cite{vonGersdorff:2005bf}, in order to obtain a
minimum at large values of the volume, one needs $c_1<0$, $c_2>0$
and
\be \label{cond}
\Big| \frac{c_2}{c_1} \Big| \gg 1 .
\ee
In a moment we will consider the concrete form of (\ref{V}) for
the $\mathbb T^6/(\mathbb Z_2\times \mathbb Z_2)$ orientifold that
we analyzed in \cite{gg1}.


\section{The $\mathbb T^6/(\mathbb Z_2\times \mathbb Z_2)$ orientifold}
\label{secZ2}

In this section we will review the known corrections to the
K\"ahler potential as discussed in the previous section in a
particular example, the $\mathbb T^6/(\mathbb Z_2\times
\mathbb Z_2)$ orientifold with D3- and D7-branes. We will see that
the  $\alpha'$ correction in the scalar potential of
the volume modulus (i.e.\ the term with coefficient $c_1$ in the previous section)
comes out with the wrong
sign for volume stabilization
in this model (more precisely for that version of
this orientifold for which the perturbative corrections were
computed in
\cite{gg1}). Nevertheless, we want to use it to give an impression
of the qualitative features and the order of magnitude of the
corrections coming from the one-loop corrections (i.e.\ of $c_2$
of the last section), the reason being that the one-loop
corrections are best understood in this particular case. We will
come back to other models in the next section.

The supersymmetric Calabi-Yau-orbifold on $\mathbb T^6/(\mathbb
Z_2\times \mathbb Z_2)$ is defined through the elements in
$\mathbb Z_2\times \mathbb Z_2=\{1,\Theta_1,\Theta_2,\Theta_3=
\Theta_1\Theta_2\}$, which all act on $\mathbb T^6=\mathbb
T^2_1\times\mathbb T_2^2\times\mathbb T_3^2$ by reflection along
four circles, $\Theta_I$ leaving $\mbb T^2_I$ invariant and
reflecting along the transverse $\mbb T^4_I$, $I=1,2,3$. The world
sheet parity projection that we consider is $\Omega'=\Omega R_6
(-1)^{F_R}$, with $R_6$ the reflection of all six circles, and $F_R$ the
right-moving fermion number. The fixed loci of $\Omega'$ and
$\Omega'\Theta_I$ define O3-planes and three sets of O7-planes.
Accordingly, there are D3-branes and three sets of D7-branes, each
of the latter wrapping one of the three $\mathbb
T_I^4$.\footnote{In
\cite{gg1} we employed the standard world sheet parity
projection with $\Omega$ instead of $\Omega'$, and thus worked
with D9- and D5-branes. Here we will translate all results into
the language of D3- and D7-branes by simple T-duality rules, as
explained in
\cite{gg1}.} Orientifold models based on $\mbb Z_2\times \mbb Z_2$ were the subject of much recent work
on moduli stabilization
\cite{Blumenhagen:2003vr,Derendinger:2004jn,Denef:2005mm,Villadoro:2005cu,Lust:2005dy}.

The untwisted moduli of the model are three K\"ahler and complex
structure moduli $\{ \rho^I, U^I\}$, the axio-dilaton $S$, and the
open string scalars. We will mostly restrict ourselves to a
representative stack of D3-branes and denote its three complex
position scalars as $A^I$, and we leave out the D7-brane scalars
and twisted moduli.\footnote{This effectively restricts us to a
subset of fluxes --- a generic flux destabilizes the orbifold
limit such that the twisted moduli receive non-vanishing
expectation values
\cite{Lust:2005bd,Denef:2005mm,DeWolfe:2005uu}.} We define the
scalars as
\beqn \label{fieldZ2}
S &=& \frac{1}{\sqrt{8\pi^2}} (C + i e^{-\Phi} ) \ , \quad
U^I ~=~ \frac{1}{G^I_{44}} ( G^I_{45} + i \sqrt{G^I} )\ , \quad
A^I ~=~  a^I_4 + U^I a^I_5\ ,
\non
\rho^I &=& \frac{1}{\sqrt{8\pi^2}} (C^I_4|_{\mbb T^4_I} + i e^{-\Phi}
\cv_{\mbb T^4_I}) + \frac{1}{8\pi} A^I \frac{A^I-\bar A^I}{U^I-\bar U^I} \
.
\eeqn
The classical (sphere plus disk level) K\"ahler potential of this
model is
\beqn \label{kpotZ2cl}
K^{(0)} = -\ln ( S-\bar S) - \sum_{I=1}^3 \ln \Big[ ( \rho^I-\bar
\rho^{I}) (U^I-\bar U^{I})
 - \frac{1}{8\pi} (A^I-\bar A^{I})^2 \Big]\ .
\eeqn
Note that the moduli space does not factorize into K\"ahler moduli
$\rho^I$ and complex structure $U^I$, as long as Im$(A^I)\neq 0$.

\begin{figure}[h]
\begin{center}
\psfrag{U1}[tc][bc][0.8][1]{Re $U$}
\psfrag{U2}[tc][bc][0.8][1]{Im $U$}
\includegraphics[width=0.5\textwidth]{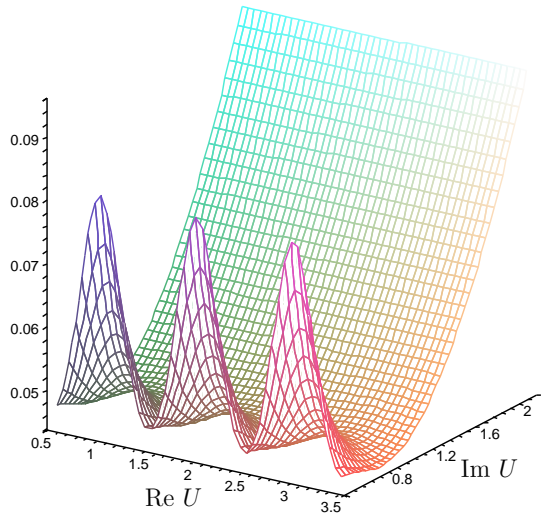}
\caption{The expression $\frac12 c\, {\mathcal E}_2^{\rm D7}(0,U)$,
where $c$ is given by $(128 \pi^6)^{-1}$ as in \cite{gg1}. The
oscillatory behavior for small imaginary part of $U$ is shown in
more detail in figure \ref{fig:E2_U1decrease}. For small real part
and large imaginary part, the function behaves as $\frac13
128\pi^4 c \times {\rm Im}(U)^2$ (see equation (B.3) in \cite{gg1}).}
\label{fig:E2_3d}
\end{center}
\end{figure}

\begin{figure}[h]
\begin{center}
\psfrag{U1}[bc][bc][0.8][1]{Re $U$}
\includegraphics[width=0.5\textwidth]{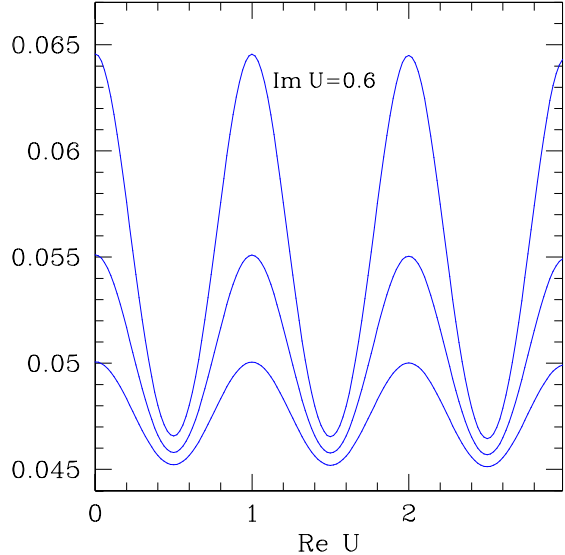}
\caption{The expression $\frac12 c\, {\mathcal E}_2^{\rm D7}(0,U)$
along Im$(U) \in\{ 0.6,0.7,0.8\}$ in figure \ref{fig:E2_3d}.}
\label{fig:E2_U1decrease}
\end{center}
\end{figure}

The one-loop correction to the K\"ahler potential is given by (see
equation (3.30) of \cite{gg1})
\beqn \label{kpotZ2}
K^{(1)} =
\frac{1}{256\pi^6} \sum_{I=1}^3 \Bigg[
 \frac{\ce^{\rm D3}_2(A^I,U^I)}{(S-\bar S)(\rho^I-\bar \rho^I)} +
 \frac{\ce^{\rm D7}_2(0,U^I)}{(\rho^J-\bar \rho^J)(\rho^K-\bar
\rho^K)}\Big|_{K\neq I\neq J} \Bigg]\ .
\eeqn
The superscripts ``D3'' and ``D7'' are used to indicate that the
two terms require the presence of the respective type of
D$p$-brane in order to be non-vanishing, since they originate from
open string diagrams with at least one boundary on these branes.
Some comments are in order. In defining the function $\ce^{{\rm
D}p}_2(A,U)$, we must ensure that a certain anomaly constraint on
the position scalars of the D3-branes is satisfied. Until now, we
have only kept a single D3-brane modulus $A$ and ignored all other
branes. However, the other scalars cannot be set to zero entirely,
because the set of all scalars $A_i$, $i$ labelling the various
stacks with gauge groups of rank $N_i$ respectively, has to obey
$\sum_i N_i A_i =0$ while $\sum_iN_i = N_{\rm D3}$,
where $N_{\rm D3}$ is the total D3-brane charge of the O3-planes.
We write this as $N_{\rm D3}=16-N_{\rm flux}$ in terms of the
effective charge $N_{\rm flux}$ carried by background 3-form flux
\cite{Giddings:2001yu}\footnote{This way of implementing
the effect of 3-form flux on the 3-brane charge in the results
of \cite{gg1} is clearly slightly heuristic, since there was no
flux considered in that calculation, and there may be additional
corrections to the K\"ahler potential once fluxes are turned on,
as we mentioned earlier.}. The simplest solution is to consider
three stacks with $A_1=-A_2=A$, $N_1=N_2=N$, and $A_3=0$,
$N_3=N_{\rm D3}-2N$. For the D7-branes we have set all scalars to
zero from the beginning and there is no modification of the
background 7-brane charge through fluxes. In this case one has
(see equation (3.27) of \cite{gg1})
\beqn \label{E2}
\ce^{\rm D3}_2(A,U) &=& 128N \big[ E_2(A,U) + E_2(-A,U) \big]
 - 8N \big[ E_2(2A,U) + E_2(-2A,U) \big] \non
&&
 + 120 (N_{\rm D3}-2N) E_2(0,U)\ ,
\non
\ce^{\rm D7}_2(0,U) &=& 1920\,  E_2(0,U)\ ,
\eeqn
with
\beqn
\label{E2AU}
E_2(A,U) = \sum_{(n,m)\neq(0,0)} \frac{{\rm Im}(U)^2}{|n+mU|^{4}}
 \exp\Big[2\pi i \frac{A(n+m\bar U)-\bar A(n+mU)}{U-\bar U}\Big] \ .
\eeqn
Note that $E_2(A,U)$ is not holomorphic.
Things get even simpler when setting
$2N=N_{\rm D3}$. In the special case when the D3-brane charge is
completely cancelled by the 3-form flux, i.e.\ $N_{\rm flux}=16$,
the first type of correction in (\ref{kpotZ2}) is absent. We show
a plot of the function $\ce^{\rm D7}_2(0,U)$ divided by $256
\pi^6$ in figures \ref{fig:E2_3d} and \ref{fig:E2_U1decrease}.

In addition to the one-loop correction to the K\"ahler potential
there is also the tree-level $\alpha'$ correction of
\cite{Becker:2002nn} here given by
\be \label{alpha}
K^{(0)}_{\alpha'} = \frac{\chi}{2} \zeta(3) {(S-\bar
S)^{3/2} \over \sqrt{\prod_{I=1}^3 (\rho^I-\bar \rho^I)}}\ .
\ee
This correction arises from sphere diagrams and thus does not
depend on the open string moduli $A^I$. This completes the set of
corrections to the K\"ahler potential known for this model, the
full expression reading
\beqn
K = K^{(0)} + K^{(0)}_{\alpha'} + K^{(1)}\ .
\eeqn
This presents a concrete realization of (\ref{Kgeneral}) with all
coefficients determined. As mentioned above, this set of
corrections may not be complete at the given order in the large
volume expansion, due to the possibility of additional $\alpha'$
corrections that could appear at tree-level onwards in 
the presence of fluxes.

The orientifold that we considered in \cite{gg1} has gauge group
$Sp(8)^4$ and is T-dual to the model discussed in
\cite{Berkooz:1996dw}. Thus, it has Hodge numbers
$(h^{(1,1)},h^{(2,1)}) = (3,51)$, cf.\
\cite{Antoniadis:1999ux,Klein:2000qw}, and the Euler number is
$\chi = -96$, the wrong sign to achieve volume stabilization
along the lines outlined at the end of section \ref{secgeneral}.
There do exist models with the right sign of the Euler number,
such as $\mbb T^6/\mathbb{Z}_6'$ which we also considered in
\cite{gg1}. Here, the K\"ahler potential also exhibits one-loop correction terms that
are suppressed by either $(\rho - \bar \rho)^{-1}$ or $(\rho -
\bar \rho)^{-2}$, such that the structure of that model is very
similar to that of $\mathbb{Z}_2\times\mbb Z_2$.\footnote{However,
in that case we did not determine some of the exact numerical
factors yet. Moreover, we cannot exclude that there are additional
terms in the K\"ahler potential proportional to $(\rho
-\bar\rho)^{-3/2}$ from additional one-loop contributions that are
not present for $\mathbb{Z}_2\times\mbb Z_2$, but that we did not
calculate in
\cite{gg1}.} Furthermore, we expect terms with the same volume
dependence to appear also in other orientifold models, some of
which would again have the right sign of the Euler
number.\footnote{An interesting candidate is the inequivalent
version of the $\mathbb{Z}_2\times\mbb Z_2$ orientifold with Hodge
numbers $(51,3)$. In the presence of world-volume gauge fluxes on
the D-branes it allows supersymmetric solutions to the tadpole
constraints (potentially with spontaneous supersymmetry breaking),
see
\cite{Blumenhagen:2003vr,Marchesano:2004xz}.}

Awaiting more detailed studies of corrections in other models, we
continue by assuming we had indeed found a model with the right
sign of the $\alpha'$ correction and whose one-loop corrections
are qualitatively captured by the formulas of the
$\mathbb{Z}_2\times\mbb Z_2$ model discussed in this section, even
though we do not have a concrete model at hand with the full set
of corrections and all the coefficients determined. In this
spirit, we will outline a situation where stabilization of the
volume modulus is possible, but fine-tuning (of the complex
structure modulus $U$) is needed to obtain sufficiently large
volume.


\section{Perturbative volume stabilization}

We now specialize to the case where all three tori are treated
on equal footing, i.e.\ we set
\be
\rho^1=\rho^2=\rho^3\equiv \rho\ , \quad U^1=U^2=U^3 \equiv U\ , \quad
A^1=A^2=A^3 \equiv A\ .
\ee
Then the K\"ahler potential becomes
\beqn \label{Kgen}
K &=& -\ln[S-\bar{S}] -3
\ln[(\rho-\bar{\rho})(U-\bar{U})-f(A,\bar{A}, U,\bar{U})] \non
&& +\alpha {(S-\bar{S})^{3/2}
\over (\rho-\bar{\rho})^{3/2}}
+ \frac{3 c}{2} {{\mathcal E}^{\rm D3}_2(A,U)
\over (S-\bar{S})(\rho-\bar{\rho})}
+ \frac{3 c}{2} {{\mathcal E}_2^{\rm D7}(0,U)
\over (\rho-\bar{\rho})^2 } \
\eeqn
where
\be \label{consts}
c={1 \over 128 \pi^6}\ , \quad \alpha={\chi \zeta(3) \over 2}\ .
\ee
We generalized the tree-level K\"ahler potential of
$\mathbb{Z}_2\times\mbb Z_2$ slightly by allowing for an arbitrary
function $f(A,\bar{A}, U,\bar{U})$ in the shift of the argument of
the logarithm. For $\mathbb{Z}_2\times\mbb Z_2$ it is
$f(A,\bar{A}, U,\bar{U}) = (8\pi)^{-1} (A-\bar A)^2$, and
therefore independent of $U$. The resulting potential
is
\beqn
{V \over e^{K}|W|^2}
 &=&  - {3\alpha \over 4}
{S_2^{3/2} \over \rho_2^{3/2}}
+\Bigg\{
{3 \over 16}
{f^2 \over U_2^2}
-{3i \over 8}
f(\partial_{U}-\partial_{\bar{U}})f
+{3 c\over 4}{\mathcal E}_2^{\rm D7}(0,U)
+{3 \over 4}\partial_{U}f\partial_{\bar{U}}
f \\
&& \hspace{-1cm}
+{3c \over 8} {1 \over S_2}\Bigg[
{1\over 2}
\left(
{{\mathcal E}_2^{\rm D3}(A,U) \over U_2}
-i(\partial_{U}-\partial_{\bar{U}}){\mathcal E}_2^{\rm D3}(A,U)
\right) f
\non
&& \hspace{1cm}
+U_2 \Big(
\partial_{U}{\mathcal E}_2^{\rm D3}(A,U)
\partial_{\bar{U}}f
+ \partial_{\bar{U}}{\mathcal E}_2^{\rm D3}(A,U)
\partial_{U}f
\Big)
\Bigg]
\non
&& \hspace{-1cm}
+ {3 c^2 \over 16} {1 \over S_2^2} \Bigg[ U_2^2 \partial_U \ce_2^{\rm D3}(A,U)
 \partial_{\bar U} \ce_2^{\rm D3}(A,U)+ \Big(\ce_2^{\rm D3}(A,U)\Big)^2 \Bigg] \Bigg\}{1 \over \rho_2^2}
+ {\mathcal O}\left({1 \over \rho_2^{5/2}}\right)\ , \nonumber
\eeqn
where $\rho_2 = {\rm Im}(\rho)$, and so on. For the case that
$f(A,\bar{A}, U,\bar{U})$ is independent of $U$, and neglecting
the terms proportional to $({\rm Im}(S){\rm Im}(\rho)^2)^{-1}$ and
$({\rm Im}(S){\rm Im}(\rho))^{-2}$ that are suppressed for large
values of Im$(S)$,\footnote{In any case, higher-genus loop
corrections may contribute at those orders.} this reduces to
\beqn \label{potential}
{V \over e^{K}|W|^2} &=&  - {3\alpha \over 4}
{S_2^{3/2} \over \rho_2^{3/2}}
+ {3\over 4} \Bigg(
{1 \over 4}
{f^2 \over U_2^2}
+c {\mathcal E}_2^{\rm D7}(0,U)
 \Bigg){1 \over \rho_2^2}
+ {\mathcal O}\left({1 \over \rho_2^{5/2}}\right)\ .
\eeqn
Some comments are in order here. It is obvious that the no-scale
structure is already  broken by the tree-level part of the
K\"ahler potential, once the argument of the logarithm is shifted
by the function $f(A,\bar{A}, U,\bar{U})$ and the factorization of
the moduli space is lost.  Furthermore, we see that we need a
positive value for $\alpha$ to obtain a negative sign for the
$\alpha'$ correction (i.e.\ $c_1$ in formula (\ref{V})). According
to (\ref{consts}) this translates into the requirement of a
positive Euler number, which is unfortunately not met by the
$\mathbb{Z}_2\times\mbb Z_2$ orientifold in the version considered
in \cite{gg1}.

As we already announced at the end of the last section, we
continue heuristically, and ask what the prospects are
for volume stabilization in a model with the right sign of the
Euler number, under the assumption that its one-loop corrections
are similar to those of the $\mathbb{Z}_2\times\mbb Z_2$
orientifold. This is the case, for instance, for the
$\mathbb{Z}_6'$ orientifold. In this way we should be able to
capture some of the qualitative features
of volume stabilization by quantum corrections.\footnote{It
would also be interesting to better understand 
issues raised by \cite{Dudas:2004nd} in this context.}

\begin{figure}[h]
\begin{center}
\includegraphics[width=0.6\textwidth]{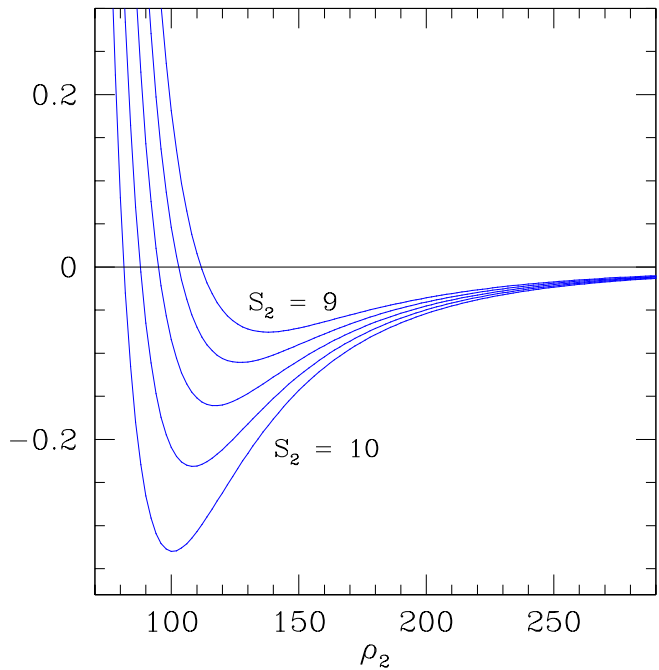}
\caption{Volume stabilization
with the potential (\ref{potential}), for a few values of Im$(S)$,
and Im$(U)=650$. The plot shows $10^{18}\times V/|W|^2$.}
\label{fig:V}
\end{center}
\end{figure}

Choosing $\alpha=50$ (corresponding to an Euler number of
the order $\chi \sim 111$) and assuming that $f(A,\bar{A},
U,\bar{U})=0$ at the minimum (this is just for technical
convenience and does not alter the result qualitatively), the
potential does have a minimum and the value of the volume at the
minimum depends on the constants $U^{(0)}$ and $S^{(0)}$, the
leading terms of the large volume expansions of $U$ and $S$ at the
minimum (we will drop the superscript $(0)$
in the following) as
\be
\label{stabvol}
\left. { \rm Im}(\rho) \right|_{\rm min} ~\sim~ \frac{(\ce_2^{\rm D7}(0,U))^2}{({\rm
Im}(S))^3}\ .
\ee
Using
the form of $E_2(A,U)$
(cf.\ (\ref{E2}) and (\ref{E2AU}))
given in (B.3) of \cite{gg1}, it follows that the
volume (\ref{stabvol}) grows roughly as
Im$(U)^4$ for large Im$(U)$ (Im$(U)\gtrsim 5$, say). Thus it is
possible to obtain relatively large values for Im$(\rho)$ at the
minimum by tuning the values where $U$ and $S$ are stabilized. In
practice it turns out that, to obtain a value of Im$(\rho) = 100$
with Im$(S)\sim 10$, one has to resort to  degenerate values for
the complex structure of about Im$(U) \sim 650$, cf.\ figure
\ref{fig:V}. Since $U$ is fixed through the flux superpotential to
leading order, we would expect typical values of the order of
$1$ to $10$, and thus consider Im$(U) \sim 650$ a significant amount of
fine-tuning.
In addition, 
one would ultimately
like to check whether higher-genus corrections might 
invalidate the genus expansion for such 
degenerate moduli values.

To summarize, the structure of the K\"ahler potential that arises
in IIB orientifold compactifications including the known $\alpha'$
and one loop corrections appears rich enough, in principle,  to
allow for purely perturbative stabilization of the volume modulus.
The necessary properties for this to work, in the case we
consider, are a positive Euler number and a tuning of the complex
structure to large values. We have not been able to present
a fully explicit model, but the $\mbb Z_6'$ orientifold comes
close. Since it is not excluded that there exist additional
corrections that cannot be neglected in the large volume and small
coupling expansion, as was discussed above, care must be exercised
in interpreting these results.


\begin{center}
{\bf Acknowledgements}
\end{center}
\vspace{-.3cm}

We would like to thank Bogdan Florea, 
Arthur Hebecker, Jan Louis, Stephan
Stieberger, and Tom Taylor for helpful advice and inspiring questions,
during discussions and email correspendence. M.B.\ was supported
by the Wenner-Gren Foundations, and M.H.\ by the German Science
Foundation (DFG). Moreover, the research of M.B.\ and M.H.\ was
supported in part by the National Science Foundation under Grant
No. PHY99-07949. The work of B.K.~was supported by the DFG, the
DAAD, and the European RTN Program MRTN-CT-2004-503369, and in
part by funds provided by the U.S. Department of Energy (D.O.E.)
under cooperative research agreement $\#$DF-FC02-94ER40818.


\end{document}